\documentclass
[prl,10pt,letterpaper,twocolumn,showpacs,bibnotes,notitlepage,final,balancelastpage,groupedaddress]{revtex4}%
\usepackage{amssymb}
\usepackage{amsmath}
\usepackage{amsfonts}
\usepackage{graphicx}%
\begin{document}
\title{Deterministic Entanglement via Molecular Dissociation in Integrated Atom Optics}
\author{Bo Zhao,$^{1}$ Zeng-Bing Chen,$^{1,2}$ Jian-Wei Pan,$^{1,2}$ J\"{o}rg
Schmiedmayer$^{1}$}
\affiliation{$^{1}$Physikalisches Institut, Universit\"{a}t Heidelberg, Philosophenweg 12,
D-69120 Heidelberg, Germany}
\affiliation{$^{2}$Hefei National Laboratory for Physical Sciences at Microscale and
Department of Modern Physics, University of Science and Technology of China,
Hefei, Anhui 230026, China}
\author{Alessio Recati,$^{3}$ Grigory E. Astrakharchik,$^{3}$ and Tommaso
Calarco$^{3,4}$} \affiliation{$^{3}$BEC-INFM, Dipartimento di
Fisica, Universit\`{a} Trento, Via Sommarive 14, I-38050 Povo,
Italy}
\affiliation{$^{4}$ITAMP, Harvard-Smithsonian Center for Astrophysics, Cambridge, MA 02138, USA}
\date{\today }

\begin{abstract}
Deterministic entanglement of neutral cold atoms can be achieved by combining
several already available techniques like the creation/dissociation of neutral
diatomic molecules, manipulating atoms with micro fabricated structures (atom
chips) and detecting single atoms with almost 100\% efficiency. Manipulating
this entanglement with integrated/linear atom optics will open a new
perspective for quantum information processing with neutral atoms.

\end{abstract}

\maketitle

The central task to most applications in quantum information processing (QIP)
and fundamental tests of quantum mechanics is the manipulation of
entanglement. Currently, the most widely used reliable source of bi-particle
entanglement is the polarization entanglement of photons, which is created via
parametric down-conversion in a nonlinear optical crystal \cite{Kwiat}. The down-conversion sources
are probabilistic, which is the major deficiency of current entangled-photon sources.
The usefulness of the photon sources also suffers from the drawbacks of
low coincidence count rate and low detection efficiency of single-photon detectors.
Entanglement creation for neutral atoms has also been proposed, which requires
controllable interactions \cite{BEC-Nature,Jaksch} or four-wave-mixing-type atomic collision
processes \cite{Yu-Niu,Aspect}.

In this Letter we present a scheme for deterministic generation and detection
of entanglement of neutral atoms. Our
scheme integrates several currently available technologies on detecting single
atoms with almost 100\% efficiency
\cite{SingAtomMOT,SingAtomCamera}, molecular
Bose-Einstein condensates \cite{mole-BEC,mole-fermi}, dissociation of
diatomic molecules \cite{Jin,Rempe,disso-Na}, and manipulating, trapping and
guiding matter waves with microfabricated structures \cite{AO}. The atom
entanglement can then be manipulated by \textquotedblleft linear atom optics
elements\textquotedblright\ \cite{MeystreBook}, which can be integrated on the Atom Chips
like atomic beam splitters (BS) \cite{atom-bs,double-well-BS}, phase shifters and
interferometers \cite{multimode-bs}. For entanglement creation we exploit the
perfect correlations inherent in an appropriately prepared two-atom state and
as such, no controlled interaction is demanded; the entanglement can be
created in either path or internal (\textquotedblleft spin\textquotedblright)
(or in both path and spin) degrees of freedom of atoms, depending on one's
demand.

Let us start by considering the free-space decay of a two-atom
system (e.g., diatomic molecule) with zero total momentum. In the
spirit of the original EPR protocol, the two atoms will freely
propagate along correlated directions due to momentum
conservation: If one of the two atoms leaves along a specific
direction, say from the left side $a_{1}$ and with momentum
$\mathbf{k}_{a}$, the remaining atom will certainly leave along
the corresponding direction $a_{2}$ (opposite to $a_{1}$) and with
momentum $-\mathbf{k}_{a}$. However the decay in free space leads
to freely propagating atoms along many pairs of correlated
directions such that the probability for the two atoms moving
along any specified pair is small.

Fortunately, one can overcome this drawback with
the help of integrated, miniaturized atom optical devices on atom chips
\cite{AO} based on microfabricated guiding structures using current carrying
wires \cite{chip}, electric charged microstructures \cite{kruger03}, and RF induced
adiabatic potentials \cite{double-well-BS}. By
restricting the decay to a limited phase space given by the atom optical
microstructure one can reduce the available decay modes significantly to only
the few desired modes. If there are only two directions the two atoms will
move along, one can deterministically obtain the path-entangled state
\begin{equation}
\left\vert \Phi\right\rangle _{path}=\alpha\left\vert a_{1}a_{2}\right\rangle
+\beta\left\vert b_{1}b_{2}\right\rangle , \label{abab}%
\end{equation}
where $\left\vert \alpha\right\vert ^{2}+\left\vert
\beta\right\vert ^{2}=1$; $\left\vert a\right\rangle $\ and
$\left\vert b\right\rangle $ are two orthonormal spatial states of
atoms. The two atoms are in a coherent superposition; the
probability amplitude $\alpha$\ ($\beta$)\ determines the
probability for the two atoms to move along the correlated
directions $a_{1}$ and $a_{2}$ ($b_{1}$ and $b_{2}$). We mention
that the free-space dissociation of molecules may lead to
continuous-variable entanglement and squeezing \cite{EPR-mole}.

A schematic drawing of the setup for our entanglement creation protocol is
shown in Fig. 1(a). In such an experimental arrangement, a diatomic molecule
is guided into a molecule BS, which can split the molecule into either
path $a$ or path $b$. In each of the one-dimensional (1D) guide, the molecule
can be dissociated (for details, see below) into
correlated atoms, which will propagate along the two pairs of correlated directions.

If the released decay energy for each atom is smaller than the transverse
level spacing in the guides, the decay can only occur in the lowest energy
state of the transverse modes, and is restricted into only one mode per path.
In this case each two-atom correlated decay leads to an atom pair
entangled in the specified paths (spatial modes). By contrast, the
parametric down-conversion for photonic entanglement creation is stochastic,
populates many modes and has a very low efficiency.

If one uses a 50-50 molecule BS in Fig. 1(a), then $\alpha$ and $\beta$ in state
$\left\vert \Phi\right\rangle _{path}$ [Eq.~(\ref{abab})] satisfy $\left\vert
\alpha\right\vert ^{2}=\left\vert \beta\right\vert ^{2}=\frac{1}{2}$ and the
two-atom state $\left\vert \Phi\right\rangle _{path}$ is a maximally entangled
state $\left\vert \Phi\right\rangle _{path}^{ME}$ for spatial modes. For
definiteness, in the following we assume one of the Bell states $\left\vert
\Phi^{-}\right\rangle _{path}\equiv\frac{1}{\sqrt{2}}(\left\vert a_{1}%
a_{2}\right\rangle -\left\vert b_{1}b_{2}\right\rangle )$\ has been
successfully created.%
\begin{figure}
[ptb]
\begin{center}
\includegraphics[
height=3cm, width=7cm
]%
{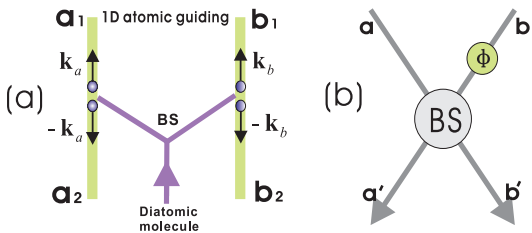}%
\caption{Generation and verification of two-atom entanglement. (a) Correlated
decay of a diatomic molecule which can be dissociated along either path $a$ or
path $b$. (b) Combining the path states of the dissociated atoms on the 50-50
atomic BS and applying phases to verify the entanglement.}%
\end{center}
\end{figure}

After the above protocol is finished, one has to verify if the two-atom
entanglement is successfully generated. To this end, one can combine the modes
$a_{1}$ and $b_{1}$ ($a_{2}$ and $b_{2}$) on the 50-50 atomic BS (see Fig. 1(b)).
Adding a phase shift $\phi$ in one of the two branches, e.g., by adjusting the depth of the guiding potentials
adiabatically \cite{multimode-bs,Yu-Niu}, the Bell
state $\left\vert \Phi^{-}\right\rangle _{path}$\ generated above becomes
$\left\vert \Phi^{\prime}\right\rangle _{path}=\frac{1}{2\sqrt{2}}[e^{i\phi
}(i\left\vert a_{1}^{\prime}\right\rangle +\left\vert b_{1}^{\prime
}\right\rangle )(i\left\vert a_{2}^{\prime}\right\rangle +\left\vert
b_{2}^{\prime}\right\rangle )-(\left\vert a_{1}^{\prime}\right\rangle
+i\left\vert b_{1}^{\prime}\right\rangle )(\left\vert a_{2}^{\prime
}\right\rangle +i\left\vert b_{2}^{\prime}\right\rangle )]$ at the outgoing
ports. Thus, the probabilities of the coincidence detections of single atoms
at any pair of the outgoing ports $1^{\prime}$ ($a_{1}^{\prime}$%
,$b_{1}^{\prime}$) and $2^{\prime}$ ($a_{2}^{\prime}$,$b_{2}^{\prime}$) are
$C_{1^{\prime}2^{\prime}}(\phi)=\frac{1}{8}\left\vert 1\pm e^{i\phi
}\right\vert ^{2}=\frac{1}{4}[1\pm\cos\phi]$ satisfying $C_{a_{1}^{\prime
}a_{2}^{\prime}}(\phi)+C_{b_{1}^{\prime}b_{2}^{\prime}}(\phi)+C_{a_{1}%
^{\prime}b_{2}^{\prime}}(\phi)+C_{b_{1}^{\prime}a_{2}^{\prime}}(\phi)=1$. Then
by observing two-atom interference fringes one can verify the successful
generation of the desired atom entanglement as the interference fringe will be
vanishing if the two probability amplitudes in $\left\vert \Phi^{-}%
\right\rangle _{path}$ are not in a coherent superposition.

The entangled pair is created deterministically, since there are
only two correlated decay channels. It is easy to see that the
fidelity of the two atom path-entanglement depends on only two
factors: the precision of linear atom optical devices, i.e., the
BS, which coherently manipulates the external state of a single
molecule; and the efficiency of the dissociation of the molecule
into two separated atoms, which determines whether we will find
the two atoms in the same outport. We will show below that one can
get a path entangled pair with fidelity of up to 97\% within our protocol using
realistic parameters.

One can also deterministically entangle internal states of two
neutral fermionic atoms by the above mechanism. Suppose one prepares a ground-state
bosonic molecule (of total spin 0) built from two bound fermionic atoms
of spin $\frac{1}{2}$. The spin\ states may correspond to, e.g., two Zeeman
sub-levels of a spin-$\frac{1}{2}$ hyperfine ground states: $\left\vert
F,m_{F}\right\rangle =\left\vert \frac{1}{2},\pm\frac{1}{2}\right\rangle $.
Such a system can easily be implemented using $^{6}$Li$_{2}$ molecules formed
from two fermionic $^{6}$Li atoms \cite{Li6Mol}. After dissociation the two fermions
will propagate along a single guide with correlations in the relative momentum
and their internal state will be in a maximally entangled spin state, e.g.,
$\left\vert \Phi^{-}\right\rangle _{spin}\equiv\frac{1}{\sqrt{2}}(\left\vert
\uparrow_{1}\downarrow_{2}\right\rangle -\left\vert \downarrow_{1}\uparrow
_{2}\right\rangle )$.

Detection of the entanglement is analogous to that for polarization
entanglement of photons. A projection to the $+45/-45$ bases can be
accomplished by applying a $\pi/2$ RF pulse which transforms $\left\vert
\uparrow\right\rangle \rightarrow\left\vert +45\right\rangle =\frac{1}%
{\sqrt{2}}(\left\vert \uparrow\right\rangle +\left\vert \downarrow
\right\rangle )$ and $\left\vert \downarrow\right\rangle \rightarrow\left\vert
-45\right\rangle =\frac{1}{\sqrt{2}}(\left\vert \uparrow\right\rangle
-\left\vert \downarrow\right\rangle )$. Two-particle interferometry can be
done using separated oscillatory field techniques borrowed from the
Ramsey-type interferometer \cite{RamseyIFM}.

Interestingly, by using the structure in Fig. 1(a), one can create spin-path
entanglement as following: For definiteness, suppose that the spin state is
$\left\vert \Phi^{-}\right\rangle _{spin}$ after the free decay of the two
dissociated fermionic atoms guided along either path $a_{1}$-$a_{2}$ or path
$b_{1}$-$b_{2}$. As the two possibilities for the two atoms decaying along
path $a_{1}$-$a_{2}$ or path $b_{1}$-$b_{2}$ are indistinguishable, the two
fermionic atoms must be in a state $\left\vert \Phi\right\rangle _{path}%
^{ME}\otimes\left\vert \Phi^{-}\right\rangle _{spin}$\ which is maximally
entangled both in path and in spin degrees of freedom. This kind of
\textquotedblleft double entanglement\textquotedblright\ may have important
applications, e.g., testing two-party all-versus-nothing refutation of local
realism against quantum mechanics \cite{Chen}.

Now let us consider factors that are essential for a practical implementation
of the present scheme. The required diatomic molecules can be created either
by photoassociation \cite{photoassociation} or by ramping through Feshbach
resonances \cite{Feshbach} using a time-dependent magnetic field. The Feshbach
resonance technique has recently been exploited to produce large, quantum
degenerate assemblies of diatomic molecules, starting from either an atomic
Bose-Einstein condensate \cite{mole-BEC} or a quantum degenerate gas of
fermionic atoms \cite{mole-fermi}. For our scheme we need a single
molecule. This can be prepared either by a quantum phase transition in an
optical lattice \cite{MOTT} and taking one potential well, or by extracting
one molecule using a \textquotedblleft quantum tweezer\textquotedblright%
\ \cite{tweezer} technique, which can keep the extracted molecule remaining in
the motional ground state during the operation.

After having successfully obtained the molecule in the motional
ground state, the molecule needs to be prepared in a spatial
superposition state, which can be achieved by a BS splitting it to
either path $a$ or path $b$ [ see Fig. 1(a)]. The RF induced
potentials for coherent manipulation of matter waves, which can
dynamically split a single trap into a double-well potential with
two local minima, are good candidates for BS in our protocol
\cite{adiabatic}. They allow a much better control of the external
motional trapped states than the usual BS, as demonstrated by recent
experiments on atom chips
\cite{double-well-BS}.

The correlated decay of the atom pairs, under the confined
geometry in Fig. 1(a), can be achieved either by
photo-dissociating the molecule by a bound-free transition,
detuned from the zero-energy threshold, or by sweeping the
magnetic field fast across a Feshbach resonance, as reported by
recent experiments \cite{Jin,Rempe,disso-Na}. If the sweep across
the resonance to a final field $B_{final}$ is fast enough, one can
achieve nearly mono-energetic atoms created in the dissociation.

To assess more quantitatively the impact of various factors in the
correlated decay, we consider Feshbach resonance dissociation and numerically calculate the fidelity of the path
entanglement using realistic parameters. In our protocol,
the decay process takes place in one waveguide, which is described by binary atomic collisions
in a tight atom waveguide. Tight transverse confinement fundamentally alters the
scattering properties of two colliding atoms (with coordinates
$x_{1}$ and $x_{2}$)  \cite{quasi1D,Grisa}. In this case, low
energy single-mode scattering can be described by
\begin{equation}
i\hbar\frac{\partial\psi}{\partial t}=H_{1D}\psi\equiv\left[  -\frac{\hbar
^{2}}{2\mu}\frac{\partial^{2}}{\partial x^{2}}+g_{1D}\delta(x)\right]  \psi,
\label{ham}%
\end{equation}
where $\psi$ is the relative-motion wave function and $x\equiv x_{1}-x_{2}$
is the relative coordinate of the two atoms. Here $g_{1D}=-\frac{\hbar^{2}}{\mu a_{1D}}$ is
the 1D coupling strength, the reduced mass $\mu=m/2$ and $a_{1D}=-\frac{a_{\perp
}}{2}[\frac{a_{\perp}}{a}+\zeta(1/2)]$, with $\zeta(1/2)\approx-1.46$, is the
effective 1D scattering length; $a_{\perp}=\sqrt{\frac{\hbar}{\mu\omega
_{\perp}}}$ is the transverse confinement length and $a$ is the 3D scattering
length. Near Feshbach resonance the scattering length
varies with magnetic field $B$ as $a=a_{bg}(1-\frac{\Delta B}{B-B_{0}})$,
where $a_{bg}$ is the background scattering length, $\Delta B$ is the width
and $B_{0}$ is the resonance position. Then one can access any value of the
scattering length to study the dissociation in 1D waveguide.

\begin{figure}
\begin{center}
\includegraphics[height=4.5cm,width=6.5cm]
{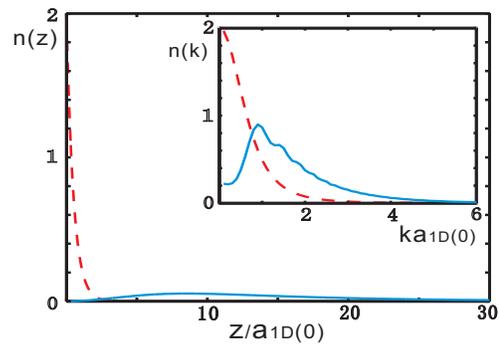}
\end{center}
\caption{The initial bound state (dashed line) and the scattered
wave-packet at a certain time after switching the interaction to a positive
value (solid line). Inset: the momentum distribution of the bound state (dashed)
and the scattered state (solid).}%
\label{evol}%
\end{figure}

We calculate the dissociation process by numerically solving the
Schr\"{o}dinger equation Eq. (\ref{ham}). In our
calculation, we choose $a_{bg}=9.2$ nm, $B_{0}=202.1$ G, $\Delta
B=7.8$ G, $\omega_{\perp}=2\pi\times69$ kHz, $a_{\perp}=85$ nm,
which are
consistent with the 1D $^{40}$K atom experiment \cite{Moritz}.
The magnetic field increases $10$ G from $208.6$ G, where $a(0)=-1.84$
nm satisfies $|a|\ll a_{\perp}$ \cite{quasi1D}, at
different magnetic field sweeping speeds. The initial relative-motion wave
function $\psi(x)=\frac{1}{\sqrt{\left\vert a_{1D}(0)\right\vert }}%
e^{-\frac{\left\vert x\right\vert }{\left\vert
a_{1D}(0)\right\vert }}$, with $a_{1D}(0)=2.04\mu$m, is the bound state of $H_{1D}$. In this process the coupling constant $g_{1D}$
goes from negative to positive value and $\psi(x)$ goes from the
bound state of $H_{1D}$ to a scattered wave packet. The typical
wave-packet evolution is depicted in Fig. \ref{evol}. It is easy
to see that the diatomic molecule is dissociated into two well
separated wave packets after sweeping the magnetic field across
the Feshbach resonance.

Taking into account the center-of-mass (CM), we can estimate the fidelity of
the path entanglement. Suppose the BS is a perfect 50-50 one,
then after dissociation the entangled state can be written as $\left\vert
\Phi^{-}\right\rangle _{path}\equiv\frac{1}{\sqrt{2}}(\left\vert\Psi_{a_{1}%
a_{2}}\right\rangle -\left\vert\Psi_{ b_{1}b_{2}}\right\rangle )$, where
$\left\vert\Psi_{a_{1}a_{2}}\right\rangle$,$\left\vert\Psi_{ b_{1}b_{2}}\right\rangle$ are
the overall wave function in the two branches respectively. The two branches are symmetric,
so we can consider the decay only in one branch, i.e., branch $a$. We have
$\Psi(x_{1},x_{2},t)=\psi(x,t)\varphi(X,t)$,
where $\psi(x,t)$ is the  relative-motion wave function and $\varphi(X,t)$
is the CM wave function in the ground
state, which can be approximated as a Gaussian wave packet. One has
$\left\vert \varphi(X,t)\right\vert ^{2}=\frac{1}{\sqrt{2\pi
}\Delta X_{t}}\exp\left[  -\frac{X^{2}}{2\left(  \Delta X_{t}\right)  ^{2}%
}\right]  $, where $\Delta X_{t}=\sqrt{\left(  \Delta X_{0}\right)
^{2}+\left(  t\Delta V_{0}\right)  ^{2}}$, with $\Delta X_{0}$\
being the initial width of the Gaussian wave packet and $\Delta
V_{0}=\frac{\hslash }{4m\Delta X_{0}}$\ the CM velocity spreading.
In this case the fidelity $F$ can be estimated by $F=1/(1+\kappa)$, where $\kappa$ is the
ratio of two second-order correlation function
$\kappa=G^{(2)}_{a_{1}a_{1}}/G^{(2)}_{a_{1}a_{2}}$, and
$G^{(2)}_{a_{1}a_{1}}$ (or $G^{(2)}_{a_{1}a_{2}}$) is just the
probability of finding the two atoms in the same output (or
opposite output). When the molecule is fully dissociated at a certain time $t_{0}$, the probability can be
approximated by
$G^{(2)}_{a_{1}a_{1}}\approx \int^{\infty}_{0}dx_{1}\int^{\infty}_{0}dx_{2}|\Psi(x_{1},x_{2},t_{0})|^{2}$
and $G^{(2)}_{a_{1}a_{2}}\approx \int^{\infty}_{0}dx_{1}\int^{0}_{-\infty}dx_{2}|\Psi(x_{1},x_{2},t_{0})|^{2}$,
where $\Psi(x_{1},x_{2},t_{0})$ is the overall two atom wave function. We numerically calculate
the fidelity for different magnetic field sweeping speeds $\dot B $ and initial widths of CM wave packet. The time $t_{0}$
is chosen 13ms so that the molecule is fully dissociated.
The results are reported in Fig. 3. We find that
the fidelity is insensitive to the sweeping speeds of magnetic field, as long as the molecule is fully dissociated.
The fidelity mainly depends on the initial width of the CM wave packet, and
one can achieve entangled pair of fidelity as high as 97\% by properly manipulating the CM wave packet.

\begin{figure}
\begin{center}
\includegraphics[height=5cm,width=7cm]
{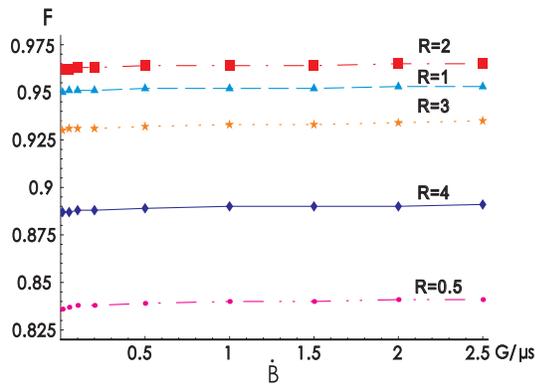}
\end{center}
\caption{Fidelity of the path entangled pair for different magnetic sweeping speeds $\dot B$
and widths of CM wave packet, with
$R\equiv \bigtriangleup X_{0}/a_{1D}(0)$.}%
\label{fid}%
\end{figure}

In summary, we have proposed a novel scheme for generating, detecting and
manipulating entanglement of neutral atoms with integrated/linear atom optics
and single-atom detection. Our scheme
entails the advantages of the usual linear optics QIP
\cite{LO-comp,LO-pure} and opens up a new avenue to QIP with neutral atoms by
means of integrated linear atom optics. The fact that the atom entanglement can be
manipulated by linear atom optics elements is interesting in its own right and
opens the exciting possibility for linear optics QIP and for
experimental test of fundamental problems in the atomic domain.

The authors would like to thank J.I. Kim for helpful discussions.
This work was supported by the European Union (ACQP, SCALA and QOQIP), the Alexander von Humboldt Foundation,
the National NSF of China, the Fok Ying Tung
Education Foundation and the CAS. We also acknowledge fundings by the European Commission under Contract
No. 509487, and
by the National Science Foundation through a grant for the Institute for the Theoretical Atomic,
Molecular and Optical Physics at Harvard University and Smithsonian Astrophysical Observatory.

\end{document}